\documentclass{article}

\usepackage[a4paper,top=2cm,bottom=2cm,left=3cm,right=3cm,marginparwidth=1.75cm]{geometry}
\usepackage[english]{babel}
\usepackage{authblk}

\usepackage{graphicx}
\usepackage{algorithm2e}
\usepackage{amsmath,amssymb}
\usepackage{dsfont}
\usepackage{multirow}
\usepackage{lineno}
\usepackage{siunitx}
\DeclareSIUnit\angstrom{\text {Å}}
\usepackage{tikz}
\usetikzlibrary{quantikz2}
\usepackage{braket}
\usepackage{graphicx}
\usepackage{ragged2e}
\usepackage{xcolor}

\usepackage[colorlinks=true,
            linkcolor=blue,
            citecolor=olive,
            filecolor=magenta,
            urlcolor=blue]{hyperref}
\usepackage{orcidlink}

\hypersetup{
    unicode=true,
    pdftoolbar=true,
    pdfmenubar=true,
    pdffitwindow=false,
    pdfstartview={FitH},
    pdftitle={Momentum microscopy of a holey nanostructure},
    pdfauthor={...}, 
    pdfsubject={PEEM},
    pdfcreator={...},
    pdfproducer={}, 
    pdfkeywords={},
    pdfnewwindow=true,
    colorlinks=true,
    linkcolor=blue,
    citecolor=blue,
    filecolor=magenta,
    urlcolor=blue
}

\usepackage[style=nature, sorting=none, backend=biber, doi=true,isbn=false,url=false,eprint=true, date=year]{biblatex}

\usepackage{csquotes}
\AtEveryBibitem{
  \clearlist{language}
}

\usepackage[figurename=Fig.]{caption}

\begin{document}

\newcommand{\WSe}{WSe\textsubscript{2} }
\newcommand{\MoSe}{MoSe\textsubscript{2} }

\title{Momentum microscopy of a holey nanostructure} 

\date{\today}

\author[1]{Kerstin\,Harland}
\author[1]{Julia\,Altenburg}
\author[2]{Xiaofei\,Wu}
\author[3]{Zsuzsanna\,Pápa}
\author[1]{Leon~David~Schwarz}
\author[1]{Lina\,Hansen}
\author[1]{Katrin\,Meier}
\author[1]{Arvid\,Klösgen}
\author[4]{Bert\,Hecht} 
\author[1]{Germann\,Hergert}
\author[5]{Péter\,Dombi}
\author[2,6]{Jer-Shing\,Huang}
\author[1]{Jan\,Vogelsang\thanks{Corresponding Author: jan.vogelsang@uol.de}}
\affil[1]{Institut für Physik, Carl von Ossietzky Universität Oldenburg, Oldenburg, Germany}
\affil[2]{Leibniz Institute of Photonic Technology, Jena, Germany}
\affil[3]{ELI ALPS, The Extreme Light Infrastructure ERIC, Szeged, Hungary}
\affil[4]{Physikalisches Institut, Universität Würzburg, Würzburg, Germany}
\affil[5]{HUN‐REN Wigner Research Center for Physics, Budapest, Hungary}
\affil[6]{Institute of Physical Chemistry and Abbe Center of Photonics, Friedrich-Schiller-Universität Jena, Jena, Germany}

\maketitle

\begin{abstract}
\noindent Momentum microscopy records the three-dimensional momentum distribution of photoelectrons emitted from a solid sample and, by utilizing sufficiently high photon energies in the ultraviolet spectral range, offers direct access to the electronic band structure of a material.
By employing a pump-probe scheme, this method provides a femtosecond view of light-matter interaction and the subsequent relaxation and transport dynamics in a broad range of material systems.
However, momentum microscopy has so far been mostly limited to flat and spatially extended surfaces.
Photoelectron emission from different three-dimensional facets, in combination with inhomogeneous electric fields around these structures, makes the analysis of the momentum distribution significantly more challenging.
Here, we extend the state-of-the-art to a substantially more complex structure and show that a careful analysis of the electron distribution recorded with a momentum microscope provides rich information about both the spatial and momentum distributions on a few-nm and few-m\AA$^{-1}$ scale, respectively.
Combined with prior knowledge of the three-dimensional structure, momentum-resolved photoelectrons can be assigned to nano-localized emission spots, permitting a new level of insight into spatio-temporal charge carrier dynamics.

\end{abstract}

\section*{Keywords}
time-resolved momentum microscopy, photoemission electron microscopy, multiphoton photoemission, electron trajectory calculations, freestanding gold film, double-nanohole antenna

\section{Introduction}

The analysis of photoelectrons emitted from a surface permits deep insights into the electronic state of the investigated sample.
While photoemission electron microscopy (PEEM) resolves the real-space emission of photoelectrons and maps local differences in the electronic configuration as a brightness contrast, momentum-resolved methods average over a selected spatial region and resolve the emission angle of the photoelectrons \cite{schmidt_time-resolved_2002, mikkelsen_photoemission_2009, dabrowski_ultrafast_2020, reutzel_probing_2024, chen_momentum_2026}.
Angle-resolved photoelectron spectroscopy (ARPES) has been one of the standard tools in surface science for a long time and is also today routinely used to resolve electronic states in solids via the detection of the kinetic energy of photoelectrons and their emission angle in one dimension \cite{feibelman_photoemission_1974, himpsel_angle-resolved_1983, reimann_subcycle_2018}.
For a complete, three-dimensional characterization, sequential recordings for different relative rotation angles of the sample and the detector are required \cite{smith_angular-resolved_1975}.
The development of delay line detectors has lifted this requirement: While a photoemission electron microscope (PEEM) could be operated in angle-resolved mode before and thus record the lateral emission angles of photoelectrons, information about the kinetic energy was missing \cite{oelsner_microspectroscopy_2001, jagutzki_multiple_2002}.
Delay line detectors, in combination with a drift tube in the microscope setup, now permit the recording of the kinetic energy of each electron in addition to its emission angle, and thus allow the experimentalist to access the three-dimensional momentum space of each electron~\cite{weber_energy-_2008}.
This new capability enabled, for example, the very recent observation of Floquet-Bloch states in graphene \cite{merboldt_observation_2025, choi_observation_2025} as well as several other exciting insights such as energy transfer in 2D-organic heterostructures \cite{bennecke_hybrid_2025}, valley-polarized Floquet-Bloch states in the transition metal dichalcogenide (TMDC) \WSe \cite{fragkos_floquet-bloch_2025}, and the structure of a Moiré exciton in a TMDC heterostructure~\cite{karni_structure_2022}.
All these works have in common that the material-vacuum interface is well-approximated by a flat plane.
This makes the analysis of electron momentum distributions easier, but also neglects the possibilities a three-dimensional structure offers in terms of imaging of charges on the nanoscale, as we will discuss below.
This can be instrumental also in nanoplasmonic photoemission experiments \cite{dombi_ultrafast_2013,teichmann_strong-field_2015,komatsu_few-cycle_2024}, e.g., for measuring and engineering nanooptical field enhancement \cite{racz_measurement_2017,budai_plasmon-plasmon_2018} for applications.

There have been few momentum-resolved studies on more complex surfaces, in particular thin nanowires \cite{krieg_exploring_2016}, nanorods \cite{lehr_momentum_2017} or conical tips, where the angular distribution of photoelectrons in strong laser fields has been investigated \cite{park_strong_2012}.
Hemispherical energy analyzers have been used in further investigations of nanostructures, but only the recorded kinetic energy was discussed in the corresponding publications~\cite{lovasz_nonadiabatic_2022, dombi_ultrafast_2013}.
In synchrotron facilities, light in the extreme ultraviolet to x-ray spectral range is focused to a nanometer spot and used for photoemission with subsequent angle-resolved detection~\cite{avila_antares_2023}.
Corresponding nano-ARPES studies reveal the local band structure of nanostructures such as nanowires \cite{krieg_exploring_2016}, but they have so far treated the data as if the photoelectrons were emitted from a flat surface.

Strongly structured samples could additionally enable the highly anticipated realization of attosecond streaking experiments on the nanoscale \cite{stockman_attosecond_2007, vogelsang_time-resolved_2024}.
This requires a method for localizing the excitation and probing of a sample to avoid detrimental space charge effects caused by the emission of multiple electrons per laser pulse~\cite{oloff_pump_2016}.
Structured, holey samples combined with illumination from the side opposite the detector are a solution, but the resulting complex field distributions and thus the challenging analysis of the photoemitted electrons, have so far hindered this development.

Here, we show that a tailored, complex field structure around a sample is not only a challenge for data analysis but also an opportunity to extract information from photoemitted electrons with unparalleled spatial selectivity.
We photoemit electrons from two different emitters and disentangle their emission in momentum space not only relative to each other but also resolve the origin of photoelectrons within an emitter with few-nanometer precision.
The combination of multi-dimensional experimental data with numerical modeling and a priori knowledge of the sample structure makes this development possible.
In particular, the excellent agreement between trajectory simulations and the experimental results shows how more complex surface structures can be used to control electrons after photoemission using static electric fields.
This might motivate a paradigm shift away from momentum microscopy solely on flat samples.

\section{Results and discussion}

\begin{figure*}[!htbp]
    \centering
    \includegraphics[width=0.4\textwidth]{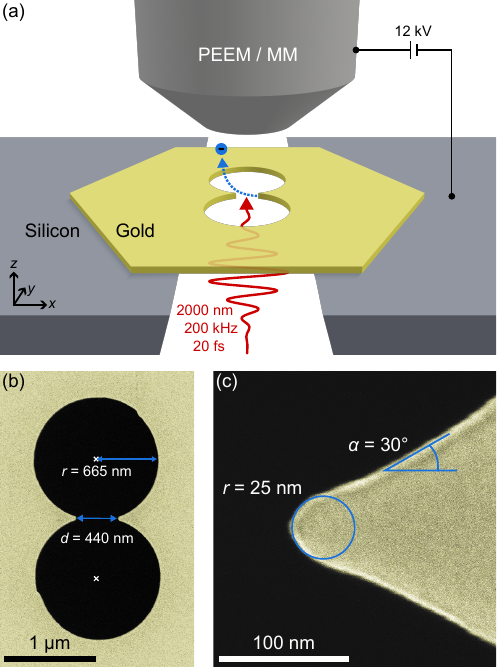}
    \caption{\justifying Experimental setup and characterization of the sample. \textbf{(a)} Sketch of the experimental setup. The nanostructured gold flake is placed on top of a silicon substrate with an aperture and illuminated from the bottom with a few-cycle light pulse (red) at a wavelength of \SI{2000}{nm}, a repetition rate of \SI{200}{kHz}, and a pulse duration of \SI{20}{fs}. The laser pulse is polarized parallel to the apices of the double-nanohole antenna and drives electron emission. The photoelectron (blue) is collected using a photoemission electron microscope operated with a potential difference of \SI{12}{kV} relative to the sample. \textbf{(b,c)} Scanning electron microscope images of the sample after the preparation. Viewed from the top, the apices of the nanostructure are characterized by a half-opening angle of \SI{30}{\degree} and a radius of curvature of \SI{25}{nm}. The thickness of the monocrystalline gold flake is \SI{55}{nm}.}
    \label{fig:setup}
\end{figure*}

We use few-cycle laser pulses in the short-wavelength infrared (SWIR) spectral range to drive nonlinear photoemission of electrons from a freestanding double-nanohole antenna and analyze the electrons using PEEM (Focus GmbH).
The experimental setup is sketched in Fig.~\ref{fig:setup}a.
The sample under investigation consists of a \SI{55}{nm} thin monocrystalline gold flake.
Using helium ion beam milling, a double-nanohole antenna with a hole radius of \SI{665}{nm} and a distance of \SI{1280}{nm} between the centers is cut into the gold flake.
An SEM image of the complete structure is shown in Fig.~\ref{fig:setup}b.
The ideal parameters of the structure, such as the radius and the distance of the two holes and the gold flake thickness, are determined by finite-difference time domain (FDTD) simulations such that it can be resonantly excited at a wavelength of \SI{2000}{nm} (see supporting information).
The determined radii and positions of the circles result in two sharp apices with radii of around \SI{25}{nm} each (see Fig.~\ref{fig:setup}c).
In the next step, the structured gold flake is placed over a \SI{50}{\micro m} by \SI{50}{\micro m} square aperture in a doped silicon substrate.
The aperture in the Si substrate gets larger toward the back side of the substrate for direct optical access to the sample on the front side, allowing a tightly focused beam to pass through the opening in the substrate.
Photoemission from the sample is driven by illumination with light pulses generated in a home-built laser system \cite{meier_multiscale_2026}.
The pulses have a central wavelength of \SI{2}{\micro m}, a repetition rate of \SI{200}{kHz}, and a pulse duration of \SI{20}{fs} (full-width at half maximum, FWHM), where the latter was characterized using the dispersion-scan technique (see the supporting information for details).
The sample is illuminated in transmission from the side opposite the detector via an off-axis parabolic mirror with a focal length of \SI{33}{mm}.
This illumination geometry, with a numerical aperture of 0.15, results in a measured laser focus diameter of \SI{7.7}{\micro m} on the sample surface. 
Additionally, it results in the simultaneous arrival of light wavefronts at all positions on a sample, as opposed to traditional illumination geometries at grazing incidence or small angles of \SI{5}{\degree} when illuminated from the sample side facing the detector.
The polarization of the few-cycle pulses is aligned parallel to the apices of the double-nanohole antenna.
The pulses provide energy to the electrons close to the gold surface in a multiphoton absorption process, eventually leading to photoemission.
The electrons are collected from the sample surface using the PEEM objective lens with an extraction voltage of \SI{12}{kV}.
Further electron optics of the PEEM project a magnified image of the emission position of each electron onto a microchannel plate~/~phosphor screen detector equipped with a complementary metal-oxide-semiconductor camera.

\subsection{Experimental results}

\begin{figure*}[!htbp]
    \centering
    \includegraphics[width=1\textwidth]{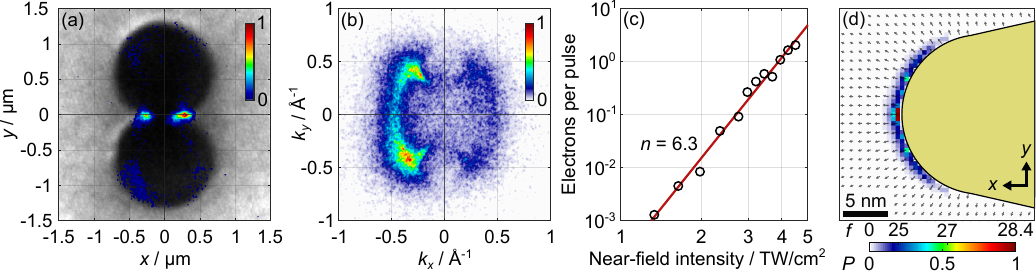}
    \caption{\justifying Nonlinear photoemission from a freestanding double-nanohole structure. \textbf{(a)} Real-space image of the nonlinear photoemission (color) from the double-nanohole apices emitted via SWIR laser pulses. The background image (gray) is acquired using ultraviolet light of a mercury lamp for photoemission. \textbf{(b)} Momentum-space image corresponding to the real-space image shown in (a). Emission from the two apices can be clearly separated, with the negative and positive $k_x$ regions corresponding to emission from the right and left apices, respectively. \textbf{(c)} Photoelectron yield as a function of the near-field intensity (black circles). A linear fit on a double-logarithmic scale provides a nonlinearity of $n = 6.3$ (red line), indicating a multiphoton photoemission process. \textbf{(d)} Numerically simulated electric field distribution in an $x$-$y$-plane \SI{7}{nm} below the top surface of the gold flake, centered around one of the two apices. The gray arrows indicate the local direction of the electric field. The emission probability $P$, which is proportional to $f^{2n}$, where $f$ is the field enhancement factor, is shown in color. The color scale represents the emission probability $P$ and is therefore nonlinear with respect to the field enhancement factor. Directly at the surface, $P$ is a measure for the expected photoelectron emission yield. Consequently, emission is primarily expected at the tip of the apex, with electrons being emitted in $x$-direction due to the local fields.
    }
    \label{fig:exp}
\end{figure*}

The resulting spatial electron distribution for laser-driven emission from the double-nanohole structure is shown in Fig.~\ref{fig:exp}a using a false color scheme.
Only two distinct emission spots are visible.
To resolve the origin of the two emission spots, the laser pulses are blocked and the sample is illuminated at grazing incidence using ultraviolet light from a mercury-vapor gas-discharge lamp (Hg-lamp), resulting in linear photoelectron emission from the surface.
The resulting spatial electron emission pattern is shown in grayscale as a background image of the laser-driven emission image.
We find that the two observed laser-driven emission spots align perfectly with the apices of the nanohole antenna.
Clearly, the antenna structure enhances the electric field strength around the two apices, localizing electron emission there.
We find that the right apex shows stronger photoemission, most likely due to a small difference in local field enhancement between the two apices.
The edge of the double-hole structure shows rather homogeneous electron emission under Hg-lamp illumination, as the high photon energy of the Hg-lamp does not resonate with the antenna structure.
Furthermore, the double-nanohole aperture is clearly discernible from the gold flake as there cannot be any electron emission from the hole itself.
During laser illumination, an additional, very weak emission signal can be observed around the edges of the whole nanostructure and, surprisingly, also inside the empty areas of the holey structure (dark blue in Fig.~\ref{fig:exp}a).
As we illuminate the back of the gold flake sample, we conclude that we observe electrons being emitted from impurities on the illumination side (bottom in Fig.~\ref{fig:setup}a).
They are then collected by the objective lens of the PEEM through the hole in the sample and thus appear to originate from the hole itself.
Generally, this needs to be considered when illuminating a holey sample from the side opposite the detector.
However, as the number of these electrons is relatively low here, we can safely ignore them in the following analysis and discussion.
We concentrate on the electrons emitted from the two sharp apices and apply a new set of voltages to the electron lenses such that, instead of the real-space emission image, the momentum distribution of the photoelectrons is projected onto the detector \cite{harland_momentum_2026}.
The resulting momentum-space distribution is shown in Fig.~\ref{fig:exp}b.

The momentum-space distribution is characterized by two distinct regions: one at negative $k_x$ values and the other one at positive $k_x$ values.
The shape of the two regions appears to be mirror-symmetric with respect to $k_x = 0$, with the left section showing a stronger signal intensity.
The intensity of the photoelectron signal in the corresponding real-space image together with considerations of the electron optics allows us to attribute the left (right) region to photoelectron emission from the right (left) apex of the nanohole structure, respectively.
Since the behavior appears to be identical for the two apices, we focus our discussion in the following on just one apex, i.e. the right apex in the real-space image with its electron momentum distribution appearing at negative $k_x$ values.
We observe that the momentum distribution is also mirror-symmetric with respect to $k_y = 0$.
This is an indication of the high quality of the sample and the symmetry of each apex, as also visible in Fig.~\ref{fig:setup}c.
Overall, the momentum distribution exhibits a C-shaped structure at $k_x = \SI{-0.45}{\per\angstrom}$ and extending up to $k_y = \SI{\pm0.6}{\per\angstrom}$.
A prominent detail is found at the ends of the C-shaped structure, where tip-like features extend toward the origin.
Yet, they do not reach the origin as almost no electrons are detected in the central low-momentum region with $\left |k \right | < \SI{0.2}{\per\angstrom}$.
This is quite surprising, as low-momentum vectors are expected in any photoemission experiment.
Part of the photoelectrons scatter incoherently before escaping the material, leading to a random distribution of momenta close to a kinetic energy of \SI{0}{eV}.
Thus, an additional effect must influence the electrons after emission before they reach the detector.

To rule out strong-field and space-charge effects interfering with the distribution, we measured the photoelectron yield per laser pulse as a function of the near-field intensity (see Fig.~\ref{fig:exp}c).
The near-field intensity is calculated from the laser power measured in front of the experimental setup, the measured focus size and pulse duration, and a field enhancement factor obtained from FDTD  (see supporting information).
A linear fit on a double-logarithmic scale reveals a multiphoton photoemission process with a nonlinearity of $n = 6.3$ in the multiphoton regime.
This further explains why laser-driven electron emission is restricted almost entirely to the two apices.
The already discussed field enhancement at the two apices leads to an enhancement of the emission rate by several orders of magnitude due to the high nonlinearity of the emission process.
Hence, emission from other areas with lower to no field enhancement is strongly suppressed.
Multiplied by the photon energy of $E_\text{ph} = \SI{0.62}{\eV}$, the measured nonlinearity gives an effective work function of $\Phi_\text{eff} = n \cdot E_\text{ph} \approx \SI{3.9}{\eV}$.
The deviation from the work function of gold ($\Phi_\text{Au} = \SI{4.8}{\eV}$ \cite{anderson_work_1959}) has been observed in photoemission experiments before \cite{zaiats_ultrafast_2026} and can most likely be explained by atomically thin surface contamination, which is not relevant for the further discussion in this work.
Additionally, the extractor field of \SI[per-mode=symbol]{5e6}{\V\per\m} of the microscope's objective lens slightly lowers the work function due to the Schottky effect, particularly when a field enhancement factor at the apices is taken into account \cite{hergert_ultra-nonlinear_2024, hergert_quenching_2025}.
When the laser field strength becomes comparable to the field binding the electrons to the atomic nuclei, it begins altering this potential periodically, leading to the closing of photoemission channels \cite{schenk_strong-field_2010, bormann_tip-enhanced_2010}.
This would lead to a deviation from the linear behavior in the double-logarithmic representation, which is not observed.
Hence, we can rule out a modification of the binding potential by a strong laser field.
Consequently, we can also exclude a relevant modification of the electron propagation after photoemission, since the laser field is not sufficiently strong to alter the electron's trajectory \cite{piglosiewicz_carrier-envelope_2014}. 

The data shown in Fig.~\ref{fig:exp}a and b were recorded at a near-field intensity of \SI[per-mode=symbol]{3.2}{\tera\W\per\centi\m\squared}.
Using the data shown in Fig.~\ref{fig:exp}c, we conclude that the images were recorded with an average electron emission rate of approximately 0.3 electrons per laser pulse. 
The number of emitted electrons per laser pulse follows a Poissonian distribution, leading to some pulses with more than one emitted electron.
However, these events account for less than \SI{15}{\percent} of the laser shots with electron emission and a comparison with literature suggests that we are still far from electron repulsion effects \cite{maklar_quantitative_2020}.

Overall, we can rule out both strong laser-field driving of the photoemitted electrons near the surface and Coulomb repulsion as the origin of the momentum distribution showing almost no signal close to zero lateral momenta.
Consequently, we postulate that the C-shaped momentum distribution must be a consequence of the sample geometry in combination with the momentum microscope optics.

To further investigate and understand this behavior, we start by examining the electric field distribution around an apex of the nanostructure.
The electric field distribution is simulated via the FDTD method using ANSYS Lumerical FDTD and the result in a $x$-$y$-plane \SI{7}{nm} below the top surface is shown in Fig.~\ref{fig:exp}d.
The arrows indicate the electric field direction during one half-cycle of the optical light field.
The local field strength is numerically determined in the simulation and raised to the power of $2n$ with the experimentally determined nonlinearity $n=6.3$.
The resulting distribution is illustrated in color in Fig.~\ref{fig:exp}d. 
Directly at the surface, this quantity is proportional to the expected photoelectron yield.
We find that the highest emission probability is localized at the left part of the structure and decreases both in the positive and the negative $y$-direction.
This behavior is qualitatively the same in all $x$-$y$-planes (see supplementary information).
Hence, most electrons will be emitted at the very tip of the apex.
After photoemission, the local electric fields in the vicinity of the apex (arrows in Fig.~\ref{fig:exp}d) slightly accelerate the electrons approximately perpendicular to the surface.
Therefore, most electrons will be accelerated primarily in the $x$-direction and will therefore only have a very small momentum in the $y$-direction.
As discussed before, this acceleration is expected to be small with the laser field strength used here and the measured in-plane momentum distribution should be rather homogeneous.
Hence, neither the observed C-shape with a strong signal for larger $k_y$ nor the tip-like features in the momentum distribution can be explained by solely looking at the emitter geometry.
Thus, we need to take the electric field distribution of the complete double nanohole in the static electric field of the objective lens of the microscope into account.

\subsection{Electron trajectory calculations}

\begin{figure*}[!htbp]
    \centering
    \includegraphics[width=0.85\textwidth]{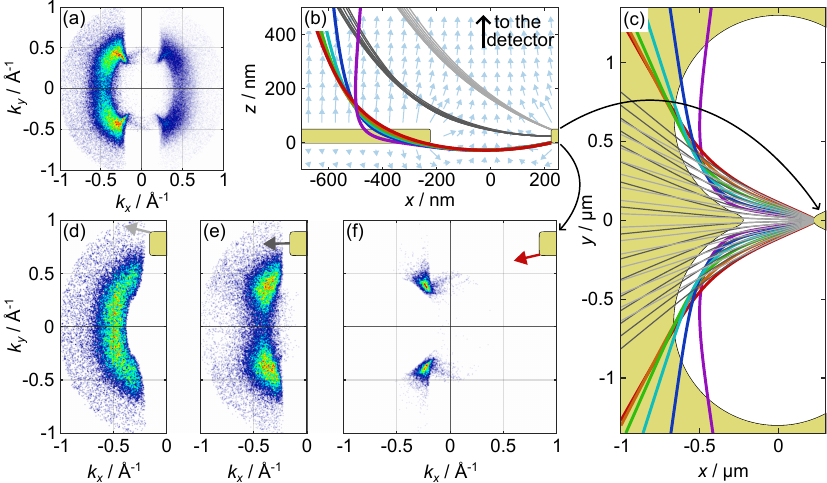}
    \caption{\justifying Simulation of the electron momentum distribution. \textbf{(a)} Result of photoelectron trajectory calculations with almost perfect agreement between experiment (see Fig.~\ref{fig:exp}b) and simulation. The distribution is simulated for one apex, which is then mirrored along $k_x = 0$ and its intensity is multiplied by 0.3. \textbf{(b)} Electron trajectories in the $x$-$z$-direction (lines) together with the local electric field directions (arrows). Electrons from three different starting planes in $z$-direction are shown, leading to distinctly different electron trajectories. \textbf{(c)} The same distribution as in (b), but shown in the $x$-$y$-direction. \textbf{(d-f)} Simulated momentum-space distributions for electrons originating from the top (d), middle (e), and bottom (f) starting planes of one apex. The starting place is indicated by a small pictogram in the top right of each panel. The origin of different momentum-space features observed in (a) and in the experiment can be clearly assigned to certain starting positions with few-nm spatial resolution.
    }
    \label{fig:sim}
\end{figure*}

Using a finite-element method, we solve Laplace's equation including the complete nanostructure and the surrounding area.
To reduce the computation space, we assume the field between the objective lens and the sample to be homogeneous after reaching a distance to the sample of several times the length of the nanostructure.
We set the boundary conditions for the electric potential such that they match the experimental conditions.
This results in a constant field strength of \SI{0.05}{V \per nm} between objective lens and sample in a distance of several micrometers to the sample.
As discussed above, we can safely ignore the laser excited near-field around the photoemitting apex.
In a next step, we propagate electrons in this electric field by numerically solving the equation of motion.
The initial conditions, including emission position, emission angle and kinetic energy, are chosen to mimic the experimental conditions.
From FDTD calculations we know that the field enhancement and thus the emission probability for an electron is higher at the top and at the bottom of the structure (see supporting information).
Thus, we tailor the emission position of the electrons to reproduce this.
At an $x$-$y$-plane several micrometers away from the sample surface, the electron momenta $k_x$ and $k_y$ are calculated from the velocities $v_x$ and $v_y$, respectively.
From all momentum vectors, a histogram is created for the $k_x$ and $k_y$ components, similar to the experimental momentum-space image.
As only the emission from one apex is simulated, the resulting momentum distribution is mirrored along $k_x = 0$ and its intensity is multiplied by 0.3 to account for the reduced emission from the left apex.
The histogram is low-pass filtered using a Gaussian kernel with a standard deviation of \SI{3.7}{\milli\per\angstrom} reproducing the experimental resolution. 
More details of the calculations can be found in the methods section below.

The resulting histogram after a complete simulation run can be found in Fig.~\ref{fig:sim}a.
Consistent with the experiment, we find a C-shaped structure, with almost no signal close to the middle of the distribution and the same triangular shapes pointing toward $k_\parallel=0$.
The excellent agreement between experiment and simulation allows us to dissect the data numerically into smaller parts to find the origin for the different features we observe.
We select three distinct groups of electrons starting, in a side view, in different $z$-positions along the apex.
Their trajectories are projected onto the $x$-$z$-plane, which is shown in Fig.~\ref{fig:sim}b.

The trajectories of the electrons starting at the top of the structure are shown in light gray, the electrons originating from the middle are plotted in dark gray and the electrons starting at the bottom, having the most dynamic behavior, are drawn in different colors.
To mimic the rounded edges of the nanostructure due to the milling process, the electrons at the top plane start with a small angle in positive $z$-direction and the electrons at the bottom start with a small angle in negative $z$-direction.
The trajectories of the latter can be better distinguished in a top view ($x$-$y$-plane), shown in Fig.~\ref{fig:sim}c.
In each of the three exemplarily selected planes, 16 electrons are started with linearly spaced starting angles limited between \SI{-0.5}{\radian} and \SI{0.5}{\radian} around the apex for illustration purposes.
Their kinetic energy is set to \SI{0.6}{\eV}, similar to the experimental conditions further discussed below.
The electrons starting in the top plane take rather straight trajectories, only marginally influenced by the field distribution around the double-nanohole antenna.
They leave the sample plane due to the extraction field of the objective lens well before getting close to the opposite apex.
This results in a rather homogeneous distribution in momentum space, shown in Fig.~\ref{fig:sim}d.
Noticeably, the distribution shows no electrons close to $k_\parallel = 0$.
By looking closer at the electrons on the $k_x$-axis for $k_y=0$, we find the explanation for this observation.
Due to the inhomogeneous field distribution around the double-nanohole structure, even electrons with a kinetic energy of \SI{0}{eV} do not exhibit a final momentum of $k_\parallel = 0$, but instead have momenta with $ k_\parallel > \SI{0.3}{\per\angstrom}$.
They gain this additional momentum $k_\parallel$ due to the parallel field components not present in traditional momentum microscopy experiments, but visible in Fig.~\ref{fig:sim}b close to the apex.
We use the experimentally measured momentum distribution on the $k_x$-axis to extract the kinetic energy distribution of the photoemitted electrons.
The experimental result is approximated by a shifted gamma distribution with a maximum around \SI{0.6}{eV} (see supporting information).
Then, each electron in the simulation is randomly assigned a kinetic energy using  this distribution.

The electrons starting in the middle plane already show a distinctly different behavior.
The extractor field is partly shielded by the metallic nanostructure such that the electric field vectors at $y = 0$ primarily point in horizontal direction ($x$) in Fig.~\ref{fig:sim}b during the first tens of nanometers after emission.
This reduces the vertical acceleration of the electrons, and thus they reach, much to our surprise, the opposite apex of the nanostructure several hundred nanometers away, despite the low starting energy in horizontal direction of just \SI{0.6}{eV}.
The electric field in the vicinity of the opposite apex alters the trajectories visibly.
The shape of the nanostructure leads to a deflection of electrons in $y$-direction.
This deflection is not random, but solely depends on the starting position of the electrons.
Electrons passing by closer to the opposite apex are deflected stronger.
Observing the corresponding dark gray trajectories shown in Fig.~\ref{fig:sim}c, we find that the ordering of the electrons remains the same, just as in the previous case.
Hence, the deflection is not strong enough to lead to a crossing of trajectories.
However, we still observe a reduction in electron density at $k_y\approx 0$ looking at the corresponding histogram shown in Fig.~\ref{fig:sim}e.
This explains the signal increase in the experiment around $k_y\approx\pm \SI{0.4}{\per\angstrom}$.
Electrons emitted closer to the middle plane of the structure, \SI{10}{nm} to \SI{20}{nm} deeper than the previously discussed electrons, reach the opposite apex and are deflected sideways in a moderate way, leading to a redistribution of detected charges in the momentum histogram from the middle outwards.

Electrons emitted from the bottom plane, due to the most direct light illumination potentially being the largest fraction in the experiment, have even further modified trajectories.
The field directly at the emitter points partially downwards, as observable in Fig.~\ref{fig:sim}b.
Hence, electrons are, although their starting velocity is almost horizontal in this model, first accelerated away from the extractor lens despite the high potential difference of \SI{12}{kV}.
The nanostructure changes the otherwise homogeneous field between sample and extractor lens and lets it point even away from the detector (arrows in Fig.~\ref{fig:sim}b).
The electrons propagate forward and reach the opposite apex at almost the same height they left their emission spot.
This leads to the strongest deflection observed so far, with some electrons even reaching the surface of the nanostructure.
We find that the number of electrons undergoing scattering is only \SI{2.7}{\percent} of the total number of electrons.
Since they do not play a significant role for the final histogram, they are also not plotted to increase the clarity of Fig.~\ref{fig:sim}b and c. 
Reflection off the surface does not result in a measurable difference or an additional feature in the momentum distribution, which is why we just remove those electrons from the calculation for improved speed.
The electrons not touching the opposite surface are nevertheless strongly deflected, as clearly observable in Fig.~\ref{fig:sim}c when looking at the colored lines.
Electrons with starting angles closer to 0 (violet trajectories) are deflected more strongly than electrons with larger starting angles (red trajectories).
This leads to the interesting situation that electrons emitted at smaller angles eventually cross the trajectory of electrons emitted at larger angles.
This behavior is only observed for electrons emitted at the lowest part of the nanostructure and thus allows us to pinpoint the origin of these electrons with few-nanometer accuracy.
Interestingly, in terms of the final momentum of the electrons, we find them grouped together in two very small spots, which can be seen in Fig.~\ref{fig:sim}f.
Hence, the deflection from the opposite apex leads to a focusing effect of the electrons in momentum space, which explains the two sharp triangular features pointing towards zero momentum.
The opposite apex acts like a sharp beam splitter for the diverging electron beam from the first apex, dissecting the beam into two parts that are focused onto two distinctly different positions on the detector. 
Thus, the starting position defines, with a few-nm sensitivity, whether an electron is accelerated to the left or to the right, providing a high spatial sensitivity.

Taken together, the higher up the electrons start, the less they are influenced by the electric field close to the nanostructure.
With the assumptions we made regarding the photoemission angles, electrons detected in momentum space can be traced back to emission spots with few-nanometer to few tens of nanometer resolution.
This provides access to parts of the nanostructure surface not resolvable so far by photoemission electron microscopy.
Additionally, the spatial resolution with respect to the origin of the electrons is even better than the spatial resolution of the microscope itself when operated in real-space mode.

\section{Conclusion}

We illuminated a free-standing double-nanohole antenna with few-cycle light pulses and photoemitted electrons from two distinct emission spots in the antenna.
The electrons were analyzed in real space and momentum space using an electron lens system.
While the real-space image shows the expected localized emission from the high-field regions of the sample, the momentum-space image reveals a surprisingly complex distribution.
We exclude the influence of space charge or strong optical fields on the distribution and trace it back to the electric fields around the nanostructure due to the potential difference between the sample and the objective lens of the electron optics.
By numerically modeling the electric field and electron propagation around the structure, we quantitatively reproduce all experimentally observed features.
In particular, we find a clear relationship between the three-dimensional emission position of an electron and its appearance on the electron detector in momentum-space mode of the lens system.
Thus, we gain insight into the two emission sites that appear as two structureless spots in the spatial imaging mode.
In our case, the apex opposite to the emitter acts as an electron beam splitter, providing a clear left/right separation.
This is a highly interesting feature for light-field-driven investigations as subtle effects of the laser field can in this way lead to a measurable difference in electron distribution.
More generally, our work shows that also structures more complex than a flat surface can be well investigated in momentum space.
With a priori knowledge on the spatial composition of a sample, the momentum-space image provides information both about the angular distribution of photoelectrons and, additionally, their emission position with few-nm resolution.
Thus, with a well-designed sample, also more complex momentum-space information could be extracted from a sample with just a few-nm spatial selectivity, close to the limitations given by the uncertainty principle.

\section*{Author contributions}
JV conceived the experiment. ZP and JV designed the sample with the help of FDTD simulations performed by ZP. XW fabricated the sample under the supervision of JSH in collaboration with BH. KM and AK prepared the optical setup and JA, LDS, KH, LH and GH performed the PEEM experiment under the supervision of JV. The experimental data was analyzed by JA, KH and JV and interpreted by JA, KH, PD and JV. Trajectory calculations were performed by JA and KH. The first draft of the manuscript was written by KH and JV and all authors contributed to the discussion and final version of the manuscript.

\section*{Funding}
We acknowledge support from the ELI-ALPS project (GOP-1.1.1-12/B-2012-000 and GINOP-2.3.6-15-2015-00001), which is supported by the European Union and co-financed by the European Regional Development Fund. The work of ZP was supported by a Bolyai Research Scholarship of the Hungarian Academy of Sciences (MTA), project nr. BO/00773/24. JSH acknowledges support by the Deutsche Forschungsgemeinschaft DFG via SFB 1375 NOA (398816777; sub-project C1) and IRTG 2675 'Meta-Active' (437527638; sub-project C1). PD acknowledges support by the National Research, Development and Innovation Office of Hungary via projects KKP137373 and TKP2021-NVA-04. JV acknowledges support by the zukunft.niedersachsen program of the Niedersächsisches Ministerium für Wissenschaft und Kultur (DyNano and Stay Inspired) and the German Research Foundation DFG (462448709, Emmy Noether program).

\section{Methods}

\subsection{Sample Fabrication.}

A monocrystalline gold flake grown on glass coverslips was selected and covered with a polymethyl methacrylate (PMMA) droplet.
After 15-minute baking at \SI{100}{\degreeCelsius} on a hot plate, the gold flake was stripped off from the glass coverslip with the PMMA droplet and transferred to another glass coverslip that was coated with a chromium (Cr) layer with void areas of different sizes.
The gold flake was placed on a slightly smaller void area, such that the outer part of the gold flake was contacting the Cr layer (to avoid charging) but most of the area of the flake was contacting the glass.
The PMMA droplet was dissolved in acetone after 10-minute baking at \SI{120}{\degreeCelsius} on a hot plate.
Outlines of the nanoantennas were then milled using a helium ion beam (HIM), on the area where the gold flake was contacting the glass.
Using the same method, the patterned gold flake was stripped off from the glass coverslip with a PMMA droplet and transferred to a silicon substrate with the nanoholes aligned to the square aperture, while the inner parts of the outline of the nanoantennas were left behind on the glass coverslip.
In the last step, the PMMA droplet was dissolved in acetone after 10-minute baking at \SI{120}{\degreeCelsius} on a hot plate.

\subsection{Simulations for the field enhancement.}

We performed finite-difference time-domain (FDTD) simulations using ANSYS Lumerical FDTD to optimize the sample geometry, and to calculate the field enhancement factor and field distributions around the apices.
The simulation volume consisted of a free-standing gold thin film with two overlapping holes. 
During optimization, we varied the layer thickness, hole diameter, and overlap between the two holes to achieve the highest field concentration around the holes.
The edges of the holes were rounded in accordance with SEM images. 
The optical data for Au was taken from the literature \cite{haynes_crc_2016}. 
To accurately describe the near-field distribution around the nanostructure, spatial resolution of \SI{0.5}{nm} in the $x$- and $y$-directions and \SI{1}{nm} in the $z$-direction was used near the apices.

The structure was illuminated with a linearly polarized short pulse centered at a wavelength of \SI{2}{\micro m}. 
To determine the field distribution and the direction of the electric field around the nanostructure, the $E_x$, $E_y$ and $E_z$ field components were collected using frequency-domain field and power monitors.
This enabled the reconstruction of the full vectorial distribution of the electric field. 
Furthermore, since the amplitude of the incident light pulse was set to unity, the resulting field-distribution maps directly represent the field enhancement factor.

\subsection{Numerical trajectory calculations.}
First, the three-dimensional geometry of the sample and the surrounding area is created.
In the center of a \SI{60}{nm}-thick disk representing the gold flake, a nanohole structure is created, consisting of two intersecting circles with radii of \SI{665}{nm} and a center-to-center distance of \SI{1280}{nm}.
Then, the resulting apices are rounded to a radius of \SI{20}{nm} each.
The free space surrounding the nanohole, in which the electric field is calculated, is cylindrical with a diameter of \SI{10}{\micro m} and a length of \SI{20}{\micro m}.
The resulting three-dimensional geometry is imported into MATLAB using the partial differential equations (PDE) toolbox. 
A locally refined mesh is used near the edges of the nanohole and a coarser mesh near the boundaries of the simulation volume.
This allows the strong spatial variations of the electric field near the nanostructure to be resolved while reducing the total number of mesh elements and, consequently, the simulation time.
The boundary conditions are chosen to match the experimental conditions.
Thus, the electric potential of the surfaces corresponding to the gold flake is set to \SI{0}{V} and the potential of the upper face is set to yield a field strength of \SI{50}{V \per \micro m}.
Utilizing the finite-element method, this electrostatic problem is solved to obtain the three dimensional electric field distribution.

The number of simulated electron trajectories in Fig.~\ref{fig:sim}a is 75000, evenly distributed across 250 starting planes along the height of one apex.
These planes are distributed more densely toward the top and the bottom of the apex.
In each plane, the electrons are randomly distributed in an ellipse ($r_x = \SI{21.25}{nm}$, $r_y = \SI{23.25}{nm}$) between \SI{-1.5}{rad} and \SI{1.5}{rad} around the apex, with the highest electron density at \SI{0}{rad}.
The electrons have an initial velocity outward from the apex in the $x$- and $y$-direction.
Electrons starting at the top and the bottom also have an initial velocity in $z$-direction to account for rounded edges due to the milling process.
The initial kinetic energy $E_\text{kin}$ is distributed between \SI{0}{eV} and \SI{4}{eV} with most electrons having a kinetic energy of \SI{0.6}{eV}. 

Each electron is then propagated independently as a classical particle by solving the equation of motion using the fourth-order Runge-Kutta method.
Collisions with the surface stop the trajectory calculation, as only a small fraction of \SI{2.7}{\percent} of the electrons either hit the sample or propagate below the sample.
Most of these electrons are starting at the bottom of the sample as they propagate the closest to the other apex.
The final electron momenta are determined from the velocities in a plane where the field distribution is homogeneous and undisturbed by the nanostructure.
The resulting k-space distribution from a single apex is mirrored along $k_x = 0$ and multiplied by 0.3 to account for the reduced emission from the other apex.

\printbibliography

\end{document}